\documentclass{caosp309}

\usepackage{graphicx}

\usepackage{natbib}
\articleNo{301}
\pubyear{2020}
\volume{50}
\volnumber{3}
\firstpage{1}
\received{May 1, 2020}
\accepted{July 28, 2020}

\def\BibTeX{{\rm B\kern-.05em{\sc i\kern-.025em b}\kern-.08em
             T\kern-.1667em\lower.7ex\hbox{E}\kern-.125emX}}

\begin{document}

\title{Long term $U$$B$$V$$R$$I$ photometric and spectral monitoring of
nova KT Eri during 2009-2023}


%
%
\author{
        S.Yu.\,Shugarov\inst{1,2}
      \and
        P.Yu.\,Golysheva\inst{2} 
        \and 
        S.\,Dallaporta\inst{3}
      \and 
        U.\,Munari\inst{4}   
      }

%
\institute{
           \lomnica, \email{shugarov@ta3.sk}
         \and 
          Sternberg Astronomy Institute, 119234 Moscow, Russia\\
        \and 
         ANS Collaboration, c/o Asronomical Observatory, 36012 Asiago (VI), Italy\\
	\and
	 INAF Padova, 36012 Asiago (VI), Italy\\
           }

\date{March 8, 2003}

\maketitle

\begin{abstract}
We present a status report of our intensive and long-term $U$$B$$V$$R$$I$
photometric monitoring of nova KT Eri since its outbust in 2009.  The
old-nova in quiescence is characterized by very high excitation conditions 
(HeII 4686 being constantly the strongest emission line in optical spectra) 
and a complex-pattern photometric variability of one mag amplitude 
in which multi-periodicities (from hours to years) are mixed with 
chaotic activity of similar amplitude. Mean color and brightness levels 
are the same for pre- and post-outburst quiescence.
\end{abstract}

%

\section{The discovery and observations of KT~Eri}

KT Eri was discovered by K.  Itagaki on 2009 November 25.536 UT at 8.1mag,
already on a fast ($t_2$=6.6 days) decline from maximum
brightness reached on November 14.67 UT at unfiltered 5.4 mag
\citep{2010ATel.2558....1H}. Although several properties of KT~Eri suggest it
could be a recurrent nova, searches for previous outbursts proved (so far)
unfruitful  \citep{2012A&A...537A..34J},~ \citep{2022MNRAS.517.3864S}.  

\begin{figure}[h!]
\centerline{\includegraphics[width=0.77\textwidth,clip=]{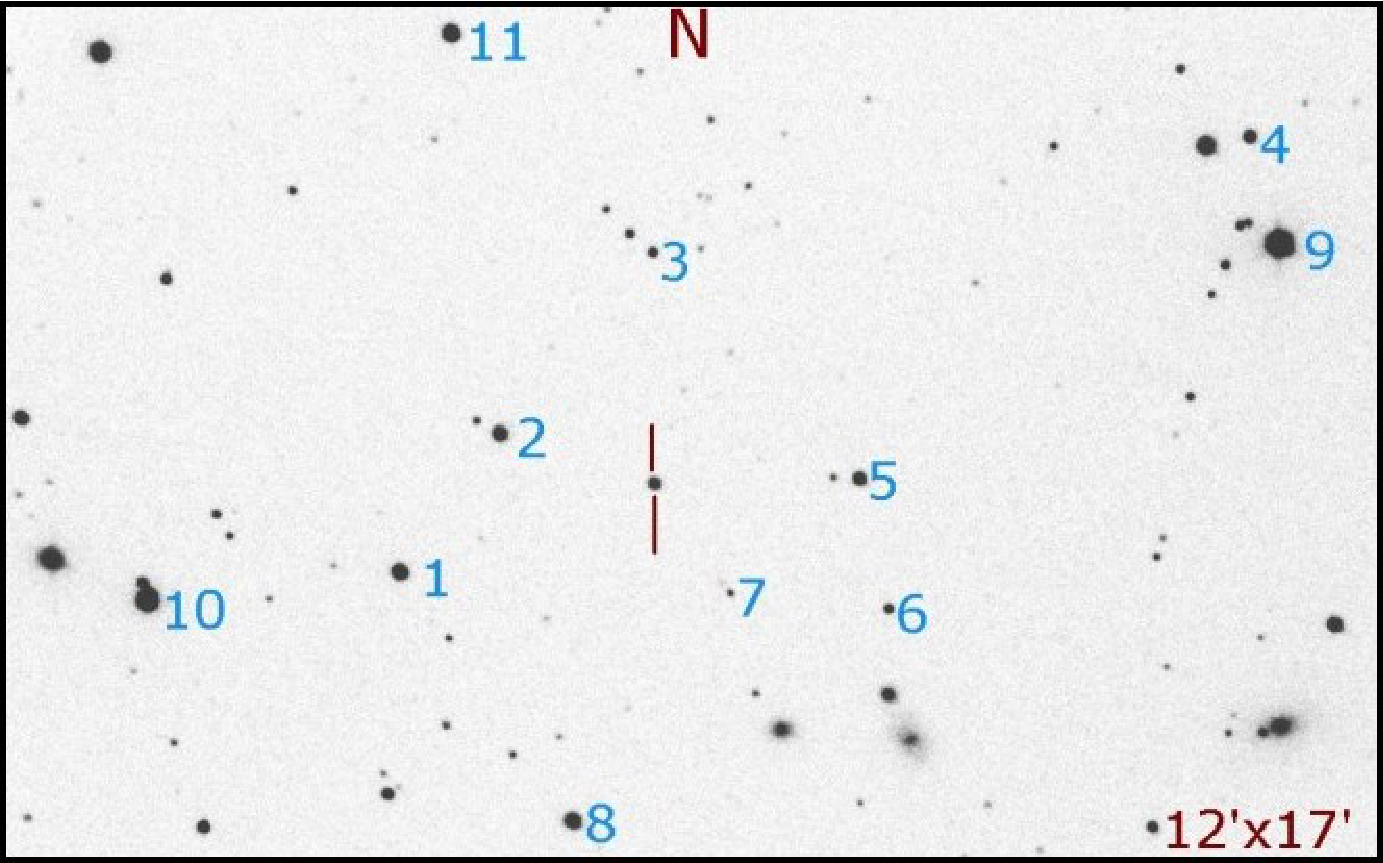}}
\caption{Identification of our photometric comparison sequence around KT~Eri.}
\vspace{-4mm}
\label{map}
\end{figure}
\begin{figure}[h]
\centerline{\includegraphics[width=1.00\textwidth,clip=]{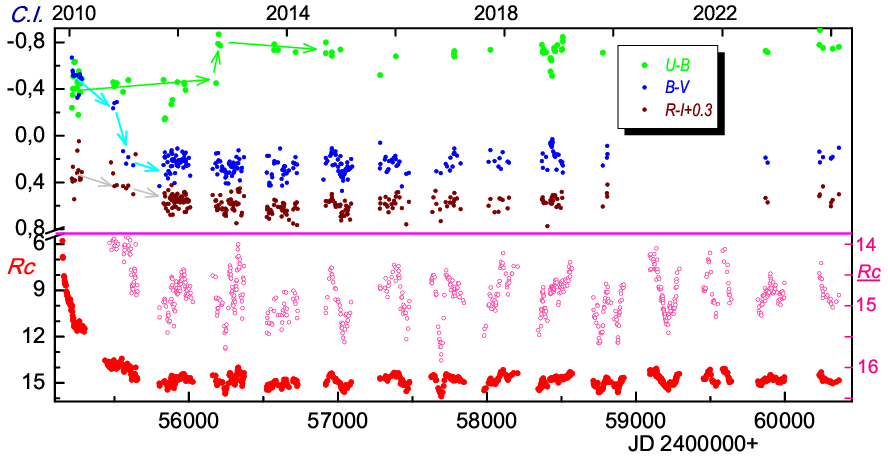}}
\caption{The overall light curve of KT Eri in $R$ (lower panel) and color
indices (upper panel) during the outburst and in quiescence.  For a better
visualization, the $R$-band lightcurve of the lower panel is plotted on both
full and expanded scales (cf. ordinates on the left and right axes).} 
\label{LC}
\end{figure}

In January 2010 we begun an intensive $U$$B$$V$$R$$I$ photometric monitoring
of KT Eri, with a broad assortment of 18--125cm telescopes located in Slovak
Republic, Crimea, and Italy.  All the photometry has been calibrated against
the same photometric sequence presented in Tab.\,\ref{t1} and Fig.\ref{map},
which has been carefully tighted to the Landolt's system of equatorial
standards \citep{2009AJ....137.4186L}. Our color- and light-curves of KT~Eri
over 2009-2024 are shown in Fig.\,\ref{LC}.

\begin{figure}[h!]
\centerline{\includegraphics[width=0.59\textwidth,clip=, angle=270]{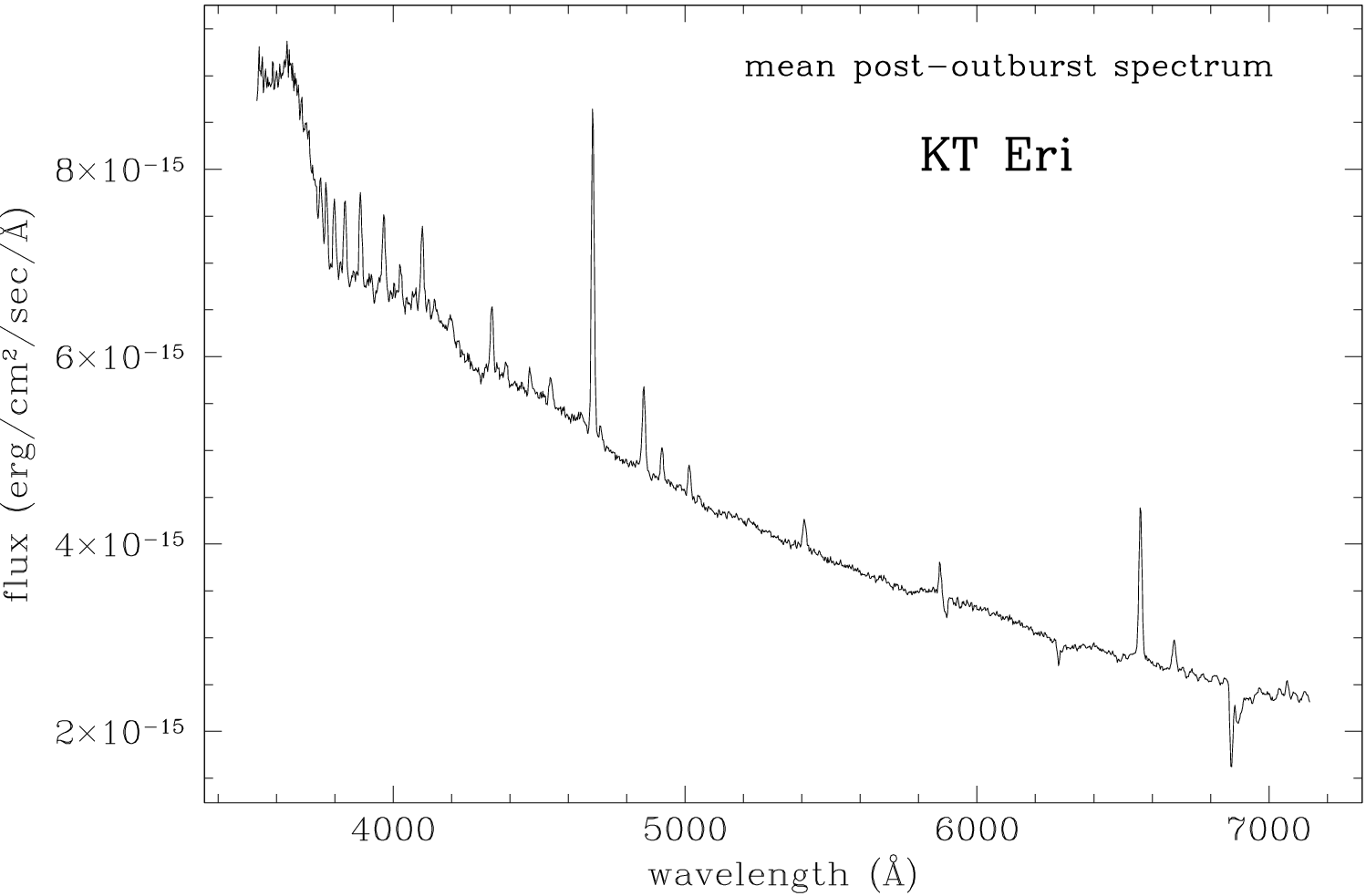}}
\vspace{3mm}
\caption{The mean optical spectrum of KT Eri during 2012--2020 yrs.}
\vspace{5mm}
\label{SP}
\end{figure}

We are also keeping KT Eri under spectroscopic monitoring with the Asiago
1.22m + B\&C in low-resolution and absolute flux modes.  The mean
post-outburst spectrum of KT Eri relative to the 2012--2020 period is
presented in Fig.\,\ref{SP}, with little changes present among different
observing epochs.  The very high excitation conditions affecting KT Eri are
readly apparent by HeII 4686 \AA\ being costantly the strongest emission line
of the whole optical range.

\begin{table}[h]
\vspace{-1mm}
\caption{Magnitudes and colors of the photometric comparison
sequence in Fig.\ref{map}.}
\vspace{-2mm}
\footnotesize
\label{t1}
\renewcommand{\tabcolsep}{0.15cm}
\begin{center}
\begin{tabular}{lcllll|lcllll}
\hline\hline
star & $V$ & $U$-$B$ & $B$-$V$ & $V$-$R_C$ & $R_C$-$I_C$&star &$V$ & $U$-$B$ & $B$-$V$ & $V$-$R_C$ & $R_C$-$I_C$ \\
\hline
1  &  12.820 & 0.81  &  0.998 & 0.575 & 0.503 &  7  & 16.345 & 0.01  & 0.565 & 0.370 & 0.334  \\
2  &  13.267 & 1.10  &  1.256 & 0.816 & 0.644 &  8  & 12.694 & 0.88  & 0.981 & 0.571 & 0.488  \\
3  &  15.197 & -0.01 &  0.540 & 0.386 & 0.388 &  9  & 10.172 & 0.07  & 0.540 & 0.37  & 0.31   \\
4  &  13.844 & 0.41  &  0.740 & 0.454 & 0.381 & 10  & 11.561 & 0.09  & 0.55  & 0.32  & 0.34   \\
5  &  13.719 & 0.93  &  0.981 & 0.562 & 0.474 & 11  & 12.420 & 0.31  & 0.590 & 0.387 & 0.336  \\
6  &  15.099 & -0.07 &  0.540 & 0.361 & 0.339 &     &        &       &       &       &        \\
\hline\hline
\vspace{-9mm}
\end{tabular}
\end{center}
\end{table}
\normalsize

\begin{figure}[h!]
\vspace{-1mm}
\centerline{\includegraphics[width=0.9\textwidth,clip=]{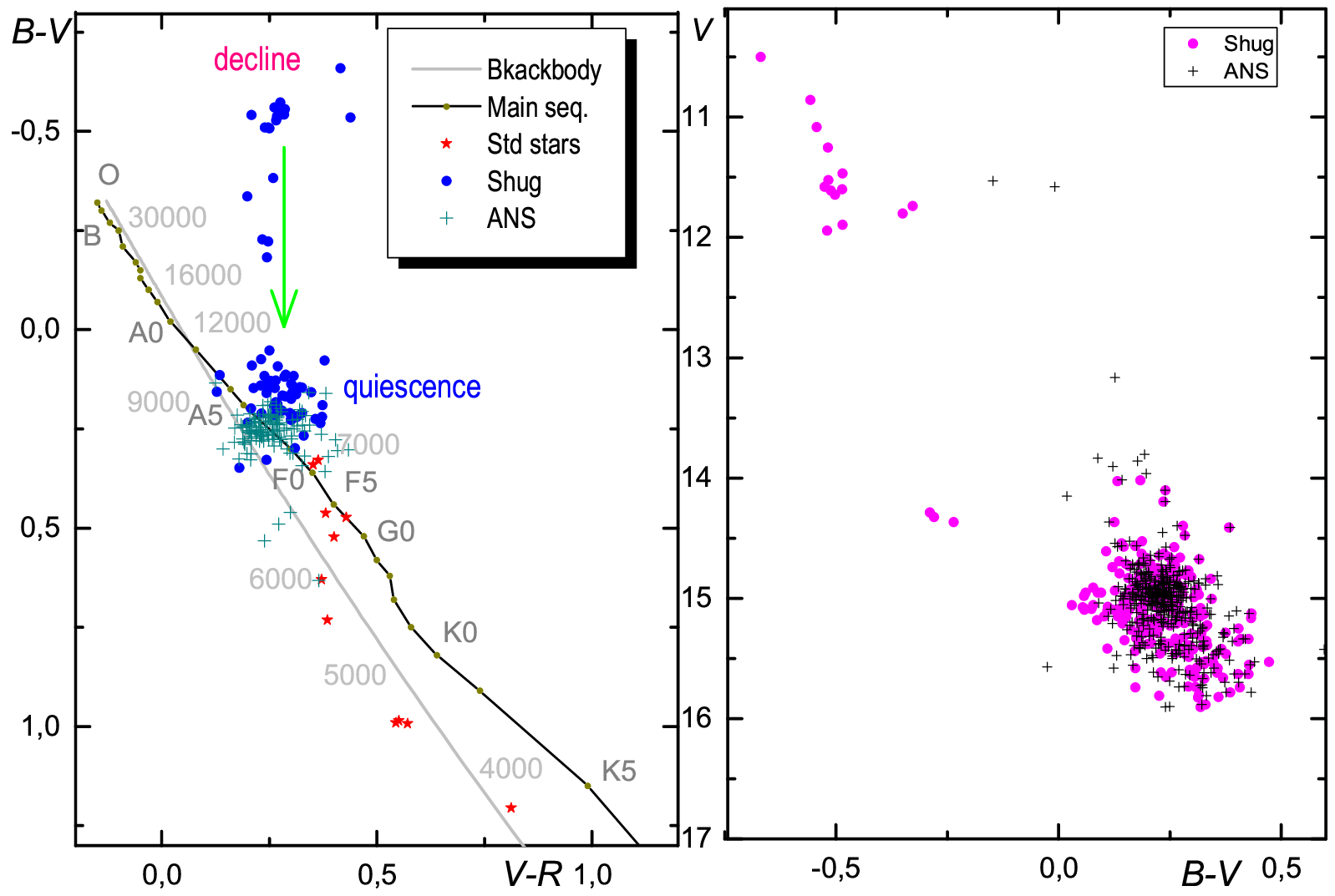}}
\vspace{-1mm}
\caption{The object's track on the two-color diagram $V$$-$$R$,
$B$$-$$V$ (left) and the $V$ vs. ($B$$-$$V$) plane (right).}
\vspace{-2mm}
\label{C4}
\end{figure}

\section{Results}

Our color- and light-curves of KT Eri are presented in Fig.~\ref{LC}.  The
decline time from 2009 maximum is derived to be $t_3$=13 days, confirming
the classification as a very fast nova.  KT Eri returned quickly to
quiescence, leveling off at the same $R$=15.3 mag characterizing its
pre-outburst brightness on Palomar plates.  The photometric post-outburst
quiescence shows the same large variability ($\sim$1 mag) inferred from
inspection of pre-outburst photographic plates \citep{2012A&A...537A..34J}. 
It is characterized by various time scales or multi-periodicities,
intricately connected and not easy to isolate because not always all present
at the same time, which are mixed with a strong chaotic activity that may
reach an amplitude of 0.7 mag over a few tens of minutes.  The broad
assortment of photometric times scales affecting KT Eri \citep[from hours to
years, eg.][]{2014NewA...27...25M,2022MNRAS.517.3864S} has already been
noted.  We plan to carry out a comprehensive analisys in search for
persistent periodicities by combining pre- and post-outburst available
data.

The star track on the $B$$-$$V$, $V$$-$$R$ planes is plotted in
Fig.~\ref{C4}.  Apart from bluer $B$$-$$V$ colors during the decline from
2009 outburst maximum, the star has remained rather stable in color during
quiescence, at an averaged $B$$-$$V$=$+$0.22, $V$$-$$R$=$+$0.25.

\acknowledgements
SS is grateful for financial support from grants APVV-20-0148, VEGA
2/0030/21 and VEGA 2/0031/22.  PG is grateful to the grants RSF-14-12-00146
and SAIA (The National Scholarship Programme of the Slovak Republic).
UM acknowledges support from INAF 2023 Minigrant funding program. 

\bibliography{demo_caosp309}

\end{document}